\documentstyle[aps,prl,multicol,epsf]{revtex} 
\begin{document}
\title{Inelastic collapse of a randomly forced particle}
\author{Stephen J. Cornell$^{1,2*}$,
Michael R. Swift$^1$, and Alan J. Bray$^1$\\} 
\address{
$^{1}$Department of Theoretical Physics,
 University of Manchester, Manchester, M13 9PL, UK.\\
$^{2}$Laboratoire de Physique Quantique, Universit\'e Paul Sabatier, 
31062 Toulouse Cedex, France.}
\date{21 April 1998}
\maketitle
\begin{abstract}
We consider a randomly forced particle moving in a finite 
region, which rebounds inelastically with coefficient of restitution $r$ 
on collision with the boundaries.  We show that there is a 
transition at a 
critical value of $r$, $r_c\equiv e^{-\pi/\sqrt{3}}$, above which the dynamics 
is ergodic 
but beneath which the particle undergoes inelastic collapse,  
coming to rest after an infinite number of collisions in a finite 
time. The value of $r_c$ is argued to be independent of the
size of the region or the presence of a viscous damping term in the
equation of motion.
\end{abstract}
\begin{multicols}{2}

Randomly accelerated particles have been used as models in many contexts,
ranging from the classic problem of
Brownian motion \cite{bm} to ecological phenomena such as
swarming \cite{grun}.
Theoretical studies have centered on first-passage problems both with
\cite{mw1}
and without \cite{burk,porra} viscous damping or in the presence of a
potential \cite{duck}.
Recently, systems of such particles that collide inelastically have
been studied in connection with granular media \cite{rev}. 
These systems are
characterized by clustering \cite{macintosh}, 
which is thought to be
a collective effect related to inelastic collapse 
of ballistically moving particles \cite{mcn}.

In this Letter we show that inelastic collapse can occur in a system
consisting of a {\sl single} particle.  
Specifically, we study a randomly forced
inelastic particle in a finite box.
If the coefficient of restitution governing the collisions with
the boundaries is less than some critical value, $r_c$, the particle
will collide with one wall an infinite number of times in a
finite time and come to rest at the wall. The value of $r_c$ is shown to be
independent of the system size or the presence of viscous damping.
This behaviour
represents a novel example of a broken ergodicity transition in a single
particle, one-dimensional stochastic system.  It also has
repercussions on the use of randomly-forced inelastic particles as
models of granular media.

Consider a particle moving with position $x$ at time $t$ within a
region of size $l$ and subject to a random force. The equation of
motion is
\begin{equation}
{d^2x\over dt^2}=\eta(t),\label{eom1}
\end{equation}
for $0<x<l$, where $\eta(t)$ is a Gaussian white noise with correlator
$\langle{\eta(t_1)\eta(t_2)}\rangle=2\delta(t_1-t_2)$ and initially
we ignore viscous damping.  
When the particle
collides with one of the walls it rebounds inelastically with
coefficient of restitution $r$, i.e., $v_f=-rv_i$, where $v_i$ and $v_f$ are
respectively 
the velocities  just before and just after the collision.  The random
force tends to increase the particle's energy, while the collisions
dissipate energy.  One might naively expect that the motion would
settle into a steady state with a well-defined average energy, except
in the pathological cases $r=0$ (where the particle simply adheres to
the first wall it hits) and $r=1$ (where there is no dissipation and
the particle's energy increases without limit).

Solutions to the Focker-Planck equation for a randomly accelerated
particle are technically formidable, even in time-independent situations
where the boundary conditions are known \cite{porra}.  In the present problem, 
the boundary conditions
must be determined self-consistently, so that the flux of particles
leaving the system with velocity $-|v|$ is balanced by a flux of
particles re-entering the system with velocity $r|v|$.  We have not
been able to obtain explicit solutions either for the time-dependent
case with specific initial conditions, nor for a putative steady
state.  
However, exact arguments about the behaviour of the system may be
formulated by mapping the motion onto that of an elastic particle.

We discuss first the semi-infinite case $l=\infty$.  The following
transformation 
\begin{equation}
x\to x'=r^{-3}x;\qquad t\to t'=r^{-2}t\label{trans1}
\end{equation}
leaves the equation of motion (\ref{eom1}) and the variance of the
noise invariant.  Moreover, 
Eqs (\ref{trans1}) imply $v\to v'=r^{-1}v$, so if we perform such a
rescaling of variables after each collision then in terms
of the new variables the motion is that of a randomly accelerated
particle which collides elastically with the wall.  We
define
\def\xbar{\overline{x}}
\def\vbar{\overline{v}}
\def\tbar{\overline{t}}
\begin{equation}
\xbar=r^{-3n(\tbar)}x; \qquad
d\tbar=r^{-2n(\tbar)}dt,\label{eombar}
\end{equation}
where $n$ is the number of collisions.
Furthermore, we can remove the wall and consider the
motion of a free particle, in which case $n$ is the number of times
the particle has crossed the line $\xbar=0$ and we should take the
absolute value of $x$ when transforming back to the original variables.
The bar denotes coordinates referring to the fictitious free particle.
We can invert Eqn~(\ref{eombar}) to give:
\begin{equation}
t=\int^{\tbar}ds\,r^{2n(s)}.\label{crux}
\end{equation}
We will see that inelastic collapse occurs when $n$ 
increases with $\tbar$ in such a way that $t$ approaches a finite limit.

To facilitate the discussion, we make a further transformation onto a
stationary Gaussian process (SGP) via the following change of
variables \cite{MaSiBrCo}:
\begin{equation} 
X=\left({3\over 2}\right)^{1/2}
{\xbar \over \tbar^{\,  3/2}};\qquad
T=\ln \tbar.\label{tr2}\end{equation}
The equation of motion may be written in the form
\begin{equation}
{d^2X\over dT^2}+2{dX\over dT}+{3\over 4}X=H(T),\label{eomX}
\end{equation}
where $H(T)$ is a Gaussian white noise with correlator $\langle
H(T_1)H(T_2)\rangle= 3 \delta(T_1-T_2)$.   The correlation function 
for $X(T)$ is
$\langle X(T_0)X(T_0+T)\rangle = (3/2)e^{-T/2} - (1/2)e^{-3T/2}$.
The average number of returns
to the origin in an interval $dT$ of a SGP with correlation function
$C(T)\sim 1-AT^2$ at small $T$ is $d\langle n\rangle=\rho dT$, where
$\rho=\sqrt{2A}/\pi$ \cite{MaSiBrCo}.
Thus
$d\langle n\rangle=\rho d\tbar/\tbar$, so
$\langle n\rangle=\rho\ln\tbar$ for large $\tbar$.  
In a naive mean-field approach, we
would
replace $n(\tbar)$ by $\langle n(\tbar)\rangle$ in
Eq.~(\ref{crux}) giving $t\approx\int^{\tbar} ds\, s^{{\sqrt{3}\over
\pi}\ln r}$.  Therefore, when $r<r_c\equiv e^{-\pi/\sqrt{3}}$,
$\lim_{\,\tbar\to\infty} t$ is {\sl finite}.  This means that after a
finite time, the particle has collided with the wall an infinite
number of times and, since
$\langle x^2\rangle = \langle r^{6n}\xbar^2\rangle$, where $\langle
\xbar^2\rangle\sim \tbar^3$, the trajectory collapses and the particle
adheres to the boundary.

We now turn to an exact calculation of the return velocity distribution
after many collisions with the boundary, 
from which further details of the collapse transition can
be extracted. In particular, we will show that the value of $r_c$
predicted by the mean field analysis is correct and that the collapse
takes place in a finite time due 
the presence of a characteristic velocity scale which decays exponentially
with the number of bounces.
The calculation
involves two steps. Firstly, we will determine the speed distribution
on the 
first return to the origin for a particle governed by eqn(\ref{eom1}) and
released from $x=0$ with velocity $v_0$. 
Secondly, we will use this
distribution as a Green's function to relate the velocity distribution
at the boundary after $n$ collisions to that after $n-1$ collisions.
By iterating the resulting
recursion relation, the full return distribution after $n$ bounces can
be generated. 

The velocity distribution on the first return to the boundary
can be calculated from the
steady state properties of the following escape problem. Particles are
injected into the region $x \ge 0$
from the origin at a constant rate and with initial velocity $v_0$.
They experience a random force, eqn(\ref{eom1}), 
and can only exit from the region at $x=0$.
The steady state flux of particles leaving with speed $v$
is proportional to the first return probability, $P(v|v_0)$, for
particles released from the origin with initial velocity $v_0$.
Calculations of this type have been carried out in the context of Kramers'
equation \cite{mw2}
and are known as albedo problems in boundary layer theory.
Here we will show that the albedo solution for the undamped, random
acceleration model with delta-function injection has a simple analytic
form which, to our knowledge, has not been noted previously.

In the steady state, the probability density function $P(x,v)$
corresponding to the Langevin equation (\ref{eom1}) 
obeys the Focker-Planck
equation
\begin{equation}
{\partial^2 P(v,x)\over\partial v^2}=v{\partial P(v,x)\over\partial x}
\label{FPE}.
\end{equation}
For the escape problem outlined above, this equation
must be solved subject to the boundary conditions
\begin{eqnarray}
P(x, v)&\to&0, \quad\hbox{$v\to \pm \infty$}\label{BC1}\\
P(x, v)&\to&0, \quad\hbox{$x\to +\infty$}\label{BC2}\\
P(0, v)&=&\delta(v-v_0),\quad\hbox{for $v>0$}\label{BC3}
\end{eqnarray}
with $P(0,v)$ for $v\le 0$ the unknown function we wish to determine.
Using separation of variables, the
most general form of the solution to eqn(\ref{FPE})
which is continuous and differentiable along the line $v=0$ and
is consistent with the boundary conditions eqn(\ref{BC1}) is
\def\Ai{{\rm Ai}}
\begin{equation}
P(x,v)= \int_{-\infty}^{\infty} d\lambda\,
e^{-\lambda x} a(\lambda) \Ai (-\lambda^{\frac{1}{3}} v),
\end{equation}
where $\Ai$ is the Airy function and $a(\lambda)$ is as yet unknown.
Using the orthogonality properties of the Airy functions over
the the interval $[-\infty,\infty]$, 
this equation can be inverted to express $a(\lambda)$ as an integral
over $P(x,v)$. Along the line $x=0$ this reduces to
\begin{equation}
\lambda^\frac{1}{3} a(\lambda) =
\Ai(-\lambda^\frac{1}{3} v_0) + \int_{-\infty}^{0} dv\,
v \Ai (-\lambda^{\frac{1}{3}} v) P(0,v),
\label{AL}
\end{equation}
where we have used the boundary condition eqn(\ref{BC3}).
The requirement that $P(x,v)\to 0$ for $x \to \infty$ 
can only be satisfied if
$a(\lambda)=0$ for all $\lambda < 0$. Imposing this condition on (\ref{AL})
results in an integral equation for the unknown part of the boundary
velocity distribution,
\begin{equation}
\Ai(-\lambda^\frac{1}{3} v_0) = \int_{0}^{\infty} dv\,
v \Ai (\lambda^{\frac{1}{3}} v) P(0,-v),
\label{INT}
\end{equation}
which must hold for all $\lambda < 0$.

In the semi-infinite system the only characteristic velocity is $v_0$, so this 
necessarily sets the scale of $P(0, -v)$.  
Using the relation $P(v|v_0)=-v P(0,-v)$ and the definition
$P(v|v_0)=\frac{1}{v_0} f(\frac{v}{v_0})$,
we proceed by making the ansatz 
\begin{equation}
f(x)= \frac{3}{2 \pi} \frac{x^\frac{3}{2}}{1+x^3},
\label{AZ}
\end{equation}
which was
motivated by a numerical simulation of the above escape problem.
By expressing the Airy function in terms of Bessel functions
and performing the integral in (\ref{INT}), 
one can verify that $P(0,-v)$ obtained from the above ansatz is indeed a
solution to (\ref{INT}) for all $\lambda < 0$ as required.

Next we wish to determine how the return velocity distribution 
evolves as the particle collides many times and is
reflected with a coefficient of restitution $r$
at each bounce. Let $P_n(v)$ be
the probability density of returning to the wall with speed $v$
after $n$ collisions. 
This distribution
obeys the recursion equation
\begin{equation}
P_{n+1}(v)= \int_0^{\infty} \frac{1}{r v^\prime}
f\left(\frac{v}{r v^\prime}\right) P_n(v^\prime) d v^\prime,
\label{RR}
\end{equation}
with $f(x)$ given by (\ref{AZ}), as particles incident with speed $v$ 
are reflected with speed $r v$. 
Eqn(\ref{RR}) may be solved by first
changing variables to $u= \ln v$ which turns the
integral into a convolution. Using standard Fourier techniques,
eqn(\ref{RR}), together with the initial condition 
$P_0(v)=\delta(v-v_0)$, can be
solved up to quadrature giving
\begin{equation}
P_n(v)=\frac{1}{(2r)^n v_0}
\int_{-\infty}^{\infty} \frac{d k}{2 \pi} 
\frac{e^{i k (\ln(v/v_0) - n \ln r)}}{[\cosh \frac{\pi k}{3}]^n}.
\label{PV}
\end{equation}

We are now in a position to discuss the collapse transition in
some detail. Firstly, one can explicitly calculate the moments of the
velocity distribution $P_n(v)$. One finds
\begin{equation}
\left\langle \left(\frac{v}{v_0}\right)^\alpha \right\rangle = \left[
\frac{r^\alpha}{2 \cos \frac{\pi}{3} (1+\alpha)} \right]^n
\end{equation}
for $-5/2 \le \alpha \le 1/2$, whilest moments 
with $\alpha$ outside this range
do not exist. The fluctuations in $v$ are thus extremely large
and, as $n$ increases, 
different moments of $v$ diverge for different values of $r$.
This is because $v$ depends exponentially on $n$, so the moments 
are sensitive 
to the rare events associated with the extremes
of the return velocity distribution. However, if one is interested in
typical trajectories, the natural variable to consider is $\ln v$
as this grows only linearly with $n$ \cite{dis}. 
Changing variables in eqn(\ref{PV}) 
to $u=\ln (v/v_0)$
and defining the normalized probability distribution 
$Q_n(u)= e^u P_n(e^u)$, one finds that in the large $n$ limit,
\begin{equation}
Q_n(u)\sim n^{-\frac{1}{2}}
\exp \left[-\frac{9}{8 n \pi^2} 
\left(u-n \ln r -\frac{\pi}{\sqrt{3}} n\right)^2\right].
\end{equation}
From the peak of the distribution
we can identify a critical value of $r$, 
$r_c = e^{-{\pi}/{\sqrt{3}}}$.
Note that this value is the same as that
predicted by the mean field analysis.
Furthermore, the distribution of $\ln v$
is sharply peaked around a value $n \ln (r/r_c)$ with fluctuations
of order ${\sqrt n}$.
There is thus a typical, characteristic velocity
which behaves like $(r/r_c)^n$. On scaling grounds, one would
expect the time intervals between collisions for $r < r_c$ also to decay
exponentially with $n$ since $v \sim t^{1/2}$. 
Consequently, the total elapsed time
after an infinite number of bounces will remain finite 
and, subsequently, the particle will remain at
rest on the boundary.

We shall now discuss the case of a particle in a finite box of size
$l$.  
The trajectory of a particle starting at $x=0$ with speed $v_0$ is the
same as for the semi-infinite system up to the instant where $x=l$ for
the first time, and may be 
obtained from a trajectory of the
equivalent SGP and the
transformations (\ref{eombar}), (\ref{tr2}).  In terms of the variable 
$X(T)$, the position of the far wall is $L=l\exp(-{3\over
2}T-3n\ln r)$.  The quantity $\ln L$ decreases
linearly in $T$, but increases by $-3\ln r$ at each return of $X$ to
zero.  Figure 1 shows the logarithm of the envelope $X_{\rm max}$ of a typical 
trajectory, defined as the largest value of $|X|$ during the interval between 
the previous and the next zero, obtained
by simulating  equation (\ref{eomX}),
together with the corresponding values of $\ln L(T)$
for three values of $r$.
Because the correlations in the SGP decay exponentially in $T$,
the intervals between zeros will have short-range
correlations.  It follows that the motion of $\ln L$ is that of a biased
random walker, biased towards the origin for $r>r_c$ and away from
the origin for $r<r_c$.  Meanwhile, the distribution of $|X|$ is
very sharply cut off at $|X|\sim 1$.
It is known that a
random walker with a bias away from the origin has a non-zero
probability of never returning to the origin \cite{feller}.
Thus, for $r<r_c$, the particle has a non-zero probability of never
reaching the far wall.  If the particle does reach the far wall, it
will then have a non-zero probability of never reaching the near wall.
The particle will therefore collapse to one
of the walls, and for each trajectory $X(T)$ the time taken for the
particle to come to rest will be finite.
For $r>r_c$ the particle
will always reach the other wall in a finite time and the process will
repeat indefinitely.
 
\begin{figure}
\narrowtext
\epsfxsize=\hsize
\epsfbox{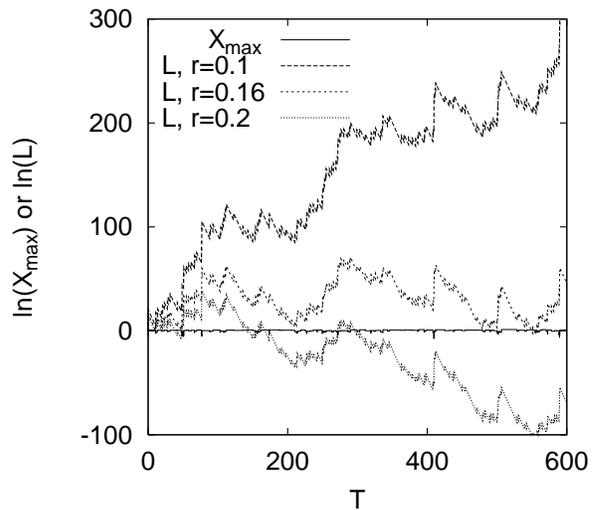}
\caption{A semilogarithmic plot of the envelope $X_{\rm max}$ (see text) 
of a typical trajectory of the SGP (Eqn.~(\protect
{\ref{eomX}})), together with the effective position of the far 
wall 
$L(T)$.  When $r<r_c=e^{-\pi/\protect{\sqrt{3}}}\approx 0.163\dots$, the 
particle 
has a non-zero chance of never reaching the wall.
}\end{figure}

There is therefore a symmetry-breaking transition between a steady state
where the particle bounces an infinite number of times from either
wall, and the collapsed state where the particle bounces a finite
number of times from one wall and an infinite number of times from the
other in a finite time.  Figure 2 shows typical trajectories for four
values of $r$, obtained from simulations of Eqn~(\ref{eomX}) and
transforming back to the real variables $x$ and $t$, showing the
collapse beneath the critical value $r_c\approx 0.163$.

\begin{figure}
\narrowtext
\epsfxsize=\hsize
\epsfbox{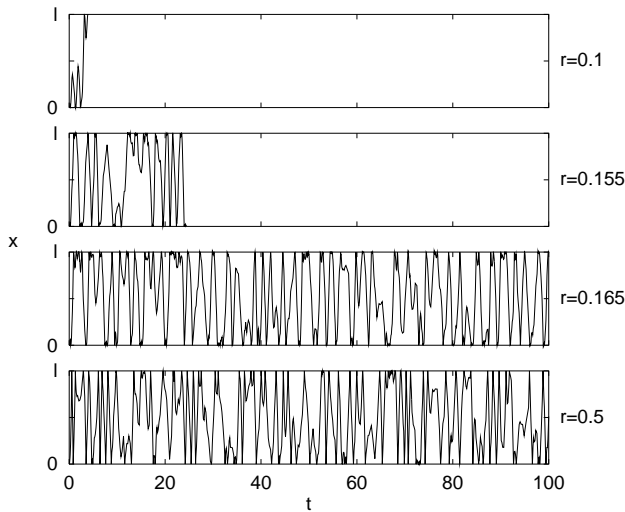}
\caption{Typical trajectories of the particle in a finite box for four values
 of $r$, showing 
inelastic collapse for $r<r_c$.}
\end{figure}

Finally we argue that a collapse
transition can occur for particles obeying Kramers'
equation of motion,
\begin{equation}
{d^2x\over dt^2} + \gamma {dx\over dt}=\eta(t).\label{eom2}
\end{equation}
This equation is often solved perturbatively by expanding
about the high viscosity limit\cite{risk}. However, if collisions
with a boundary are to be included, the effect of the inertial
term needs to be treated in a more complete way. We will again
consider a particle colliding with a single boundary
with coefficient of restitution $r$. Scaling arguments analogous
to those presented above suggest that a collapse transition
will take place for all $\gamma$ at the same value of
$r$ as in the undamped case.

We first perform a rescaling of the variables as in (\ref{trans1}).
The resulting equation of motion has the same form as (\ref{eom2})
but with a rescaled $\gamma$ given by ${\gamma^\prime} = r^2 \gamma$. After
many collisions there is an effective time dependent 
${\bar \gamma}(\tbar) = r^{2 n(\tbar)} \gamma.$
From Eqn (\ref{crux}), we sees that in the undamped problem for $r < r_c$, 
$r^{2 n({\bar t})} \to 0$ faster than $1 /{\bar t}$.
If one introduces a time dependent $\gamma $ in eqn(\ref{eom2}), 
a simple scaling analysis shows that it will be irrelevant asymptotically
if it too decays faster than $ 1/ t$. Thus, the dissipative system
will always be in the collapsed state for $r < r_c$.
Now let us assume that the dissipative system can collapse for some $r>r_c$.
As the collapsed state is approached, the velocity of the particle
goes to zero while the acceleration
remains of the order of the noise strength. Consequently the 
dissipative term $\gamma v$ will become negligible and, as we have
assumed $r > r_c$, the particle will ultimately move away from the wall.
We thus conclude that the collapse transition will occur when
$r = r_c = e^{-\pi/{\sqrt 3}}$ for any value of $\gamma$\cite{pot}.

The arguments about collapse of a single particle with a wall
can trivially be extended to two inelastic particles, in which case
the collapsed state consists of the two particles moving together.
Collapse can also be expected in a many-body system, when pairs of
particles, and then larger clusters, aggregate.  
Real systems may deviate from the ideal case studied in this Letter by
having short-range correlations in the driving force, or by the coefficient
of restitution approaching unity in the small velocity 
limit.  In these cases, one would still expect some remnant of the collapse
transition, and this should provide a method for extracting
information about the time- and velocity-scales on which such
deviations occur.  We will discuss these and other related points at greater
length elsewhere \cite{next}.

This work was supported by EPSRC and the Minist\`ere de l'Education Nationale 
et de la Recherche.

\end{multicols}

\begin{references}
\bibitem{dis} This is analogous to considering the free energy, not
the partition function, in disordered equilibrium systems.
\bibitem{pot} The same scaling approach shows that
an attractive linear potential
at the wall will also not change the value of $r_c$.
\end{references}
\end{document}